\documentclass{desyproc}
\newcommand{\jpsi}{\mathrm{J/}\psi}
\newcommand{\psip}{\psi(2\mathrm{S})}
\begin{document}
\title{Results on ultra-peripheral interactions in Pb-Pb and p-Pb collisions in ALICE}
\author{{\slshape Eugenio Scapparone$^1$ on behalf of the ALICE Collaboration,}\\[1ex]
$^1$INFN Bologna, Via Irnerio 46 40126 Bologna, Italy\\
 }
\contribID{Scapparone\_Eugenio}
\acronym{EDS'09} 
\maketitle
\begin{abstract}
Ultra-relativistic heavy ions generate strong electromagnetic fields which offer the possibility to
study $\gamma$-nucleus and $\gamma$-proton interactions at the LHC in the so called ultra-peripheral collisions~(UPC).
 Here we report ALICE results on J/$\psi$ photoproduction measured in 
Pb-Pb collisions at $\sqrt{s_{\rm NN}}$ = 2.76 TeV and in 
p-Pb collisions at $\sqrt{s_{\rm NN}}$ = 5.02 TeV.
\end{abstract}
\section{Introduction}
Collisions between heavy-ions and proton-ion at the LHC can be used to study particle production in
photonuclear interactions~\cite{ref1}. These interactions may occur in ultra-peripheral
collisions (UPC), where the impact parameter is larger than twice the hadron radius. 
The events considered here are characterized by a single $\jpsi$ meson but no
other particles being produced. 
The cross section for this process is proportional to the nuclear
gluon distribution function. In Pb-Pb collisions this measurement sheds light on the low Bjorken-$x$ region, where the nuclear gluon
shadowing is poorly known. In this talk, results on coherent photoproduction of $\jpsi$ vector
mesons measured by the ALICE Collaboration at
the LHC are presented \cite{ref8,ref9}. The $\jpsi$ is studied through its di-muon decay at forward rapidity
and both di-muon and di-electron decay at mid-rapidity. 
As far as the p-Pb collisions are concerned, the main interest is the study of the $\gamma$+ p$\rightarrow\jpsi$ + p
cross section at high center of mass energy~($W_{\gamma p}$) where gluon saturation effects
could manifest.
Coherent photoproduction is associated with a low
$p_{\rm T}$ ($<$ 100 MeV/$c$) of the final state. To take into account also the finite detector resolution, a $p_{\rm T}$
cut is used to define coherent interactions in the di-muon~(di-electron) channel for the present
analysis.
\section{The ALICE Experiment}
The ALICE detector~\cite{ref0} consists of a central barrel placed inside a large solenoid magnet (0.5 T)
covering the pseudorapidity range $|\eta |<$ 0.9, a muon spectrometer covering the range
$-4.0<y<-2.5$, and a set of smaller detectors at forward rapidity. The forward rapidity analysis is
based on the muon spectrometer. In addition, the VZERO counters and Zero-Degree
Calorimeters (ZDC) are used for triggering UPC and rejecting the contribution from hadronic
interactions, respectively.
The VZERO counters are arrays of scintillator tiles
situated on either side of the primary vertex at pseudorapidities 2.8 $<\eta <$ 5.1 (VZERO-A, on
the opposite side of the muon spectrometer) and -3.7 $<\eta <$ -1.7 (VZERO-C, on the same side as
the muon spectrometer). 
The analysis at mid-rapidity makes use of the barrel detectors (SPD, TPC and TOF), the VZERO
counters and the ZDC calorimeters. More information on the detectors used for this
analysis can be found in~\cite{ref8,ref9}.
\section{$\jpsi$ production in Pb-Pb collisions}
\subsection{Forward rapidity analysis}
The analysis at forward rapidity, is based on a sample of 3$\cdot 10^6$ events collected with a ultra-
peripheral collision dedicated trigger (FUPC) during the 2011 Pb-Pb run. The purpose of the
FUPC trigger was to select two muons in an otherwise empty detector. It is based on three
requirements: a single muon trigger above a 1 GeV/$c$ $p_{\rm T}$ threshold; at least one hit in VZERO-C; 
no hits in VZERO-A. The integrated luminosity for the data collected with this trigger
corresponds to about 55~${\rm mb}^{-1}$. 
Only
events with two opposit charged tracks in the muon spectrometer were considered. The analysis
was restricted to $\jpsi$ rapidities between -3.6 $<y<$ -2.6 and muon pseudorapidities between
$-3.7~<~\eta~<~-2.5$ to match the acceptance of the VZERO-C and avoid the edges of the
spectrometer. 
More details about the
analysis cuts are available in~\cite{ref8}. 
The invariant mass distribution
for the di-muons surviving the analysis cuts are shown in figure 1 (left). The final sample
contained 117 $\jpsi$ candidates in the invariant mass range 2.8 $< M_{inv} <$ 3.4 GeV/$c^2$. The number
of $\jpsi$ was extracted by fitting the invariant mass distribution to a Crystal Ball function, for the
signal, and an exponential, for the background from two-photon interactions. This gave an
extracted number of $\jpsi$, $N_{yield}$ = 96$\pm$12(stat)$\pm$ 6(syst). 
The transverse momentum distribution of $\jpsi$ candidates is shown in figure 1 (right). The
histograms show the expected contribution from coherent $\jpsi$ production, incoherent $\jpsi$
production, $\jpsi$ from the decay $\psip$$\rightarrow \psi$+X, and two-photon production of di-muon pairs. 
The signal region $p_{\rm T}$ $<$ 0.3 GeV/$c$ contains background contributions from
incoherent $\jpsi$ production as well as feed down from $\psip$ decay.
The method used to estimate these fractions and the associated uncertainties is explained in
detail in~\cite{ref8}. 
%
The resulting number of coherent $\jpsi$ after the above corrections is $N_{coh}^{\jpsi}$=$78\pm 10(stat)^{+7}_{-11}$(syst). 
During the 2011 Pb-Pb run, the
VZERO detector had a threshold corresponding to an energy deposit above that from a
minimum ionizing particle. This made it difficult to accurately simulate the VZERO-C trigger
efficiency for low multiplicity events. To avoid these uncertainties, the $\jpsi$ cross section was
obtained by using the number of reconstructed $\gamma\gamma\rightarrow\mu^+\mu^-$ events. The $\jpsi$ section can then
be written as
\begin{equation}
{\rm d}\sigma_{\jpsi}^{coh}/{\rm d}y= 
\frac{
N_{coh}^{\jpsi}\cdot (Acc\cdot\epsilon)_{\gamma\gamma}\cdot\sigma_{\gamma\gamma}
}
{
N_{\gamma\gamma}\cdot\Delta y\cdot BR(\jpsi\rightarrow\mu^+\mu^-)\cdot (Acc\cdot\epsilon)_{\jpsi}
}
\end{equation}
Here $(Acc\cdot \epsilon)_{\gamma\gamma}$ and 
$(Acc\cdot \epsilon)_{\jpsi}$ are the acceptance and efficiency for
two-photon and coherent $\jpsi$ events, respectively. $BR$($\jpsi\rightarrow\mu^+\mu^-)$ is the branching ratio for
the di-muon decay of the $\jpsi$ and $\Delta y$ = 1.0 is the rapidity interval. The number of 
$\gamma\gamma$ events, $N_{\gamma\gamma}$, was obtained by counting
the number of events in the invariant mass intervals 2.2 $< M_{inv} <$ 2.6 GeV/$c^2$ and 3.5 $<M_{inv}<$ 6.0 GeV/$c^2$ 
to avoid contamination from the $\jpsi$ peak. 
A cross section 
${\rm d}^{coh}_{\jpsi}/{\rm d}y= 1.00\pm0.18(stat)^{+0.24}_{-0.26}$(syst)~mb was obtained in the interval -3.6$< y <$ -2.6.
\begin{figure}[thb]
\mbox
{
\includegraphics[width=7.2cm,keepaspectratio]{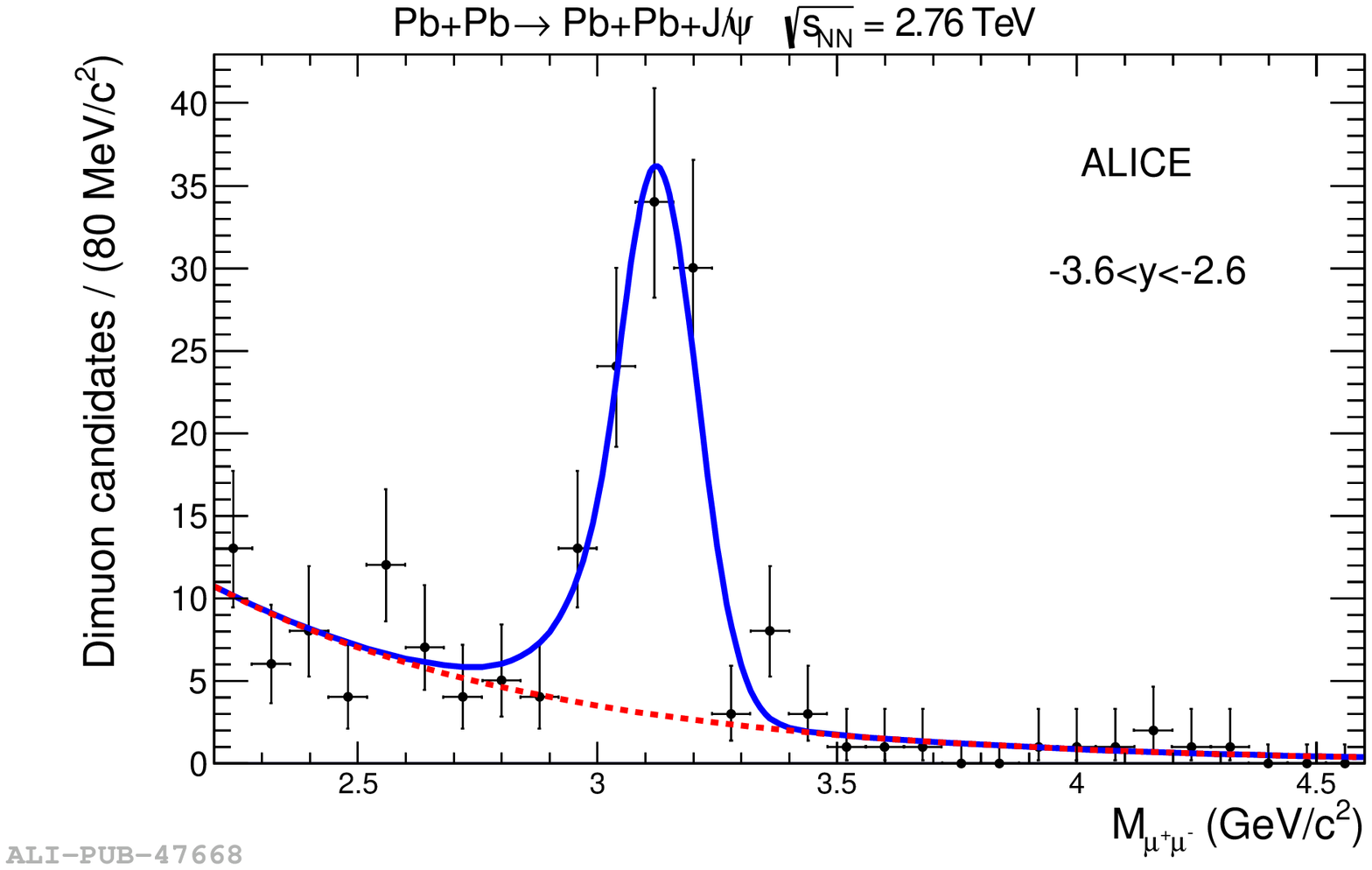}
\includegraphics[width=7.2cm,keepaspectratio]{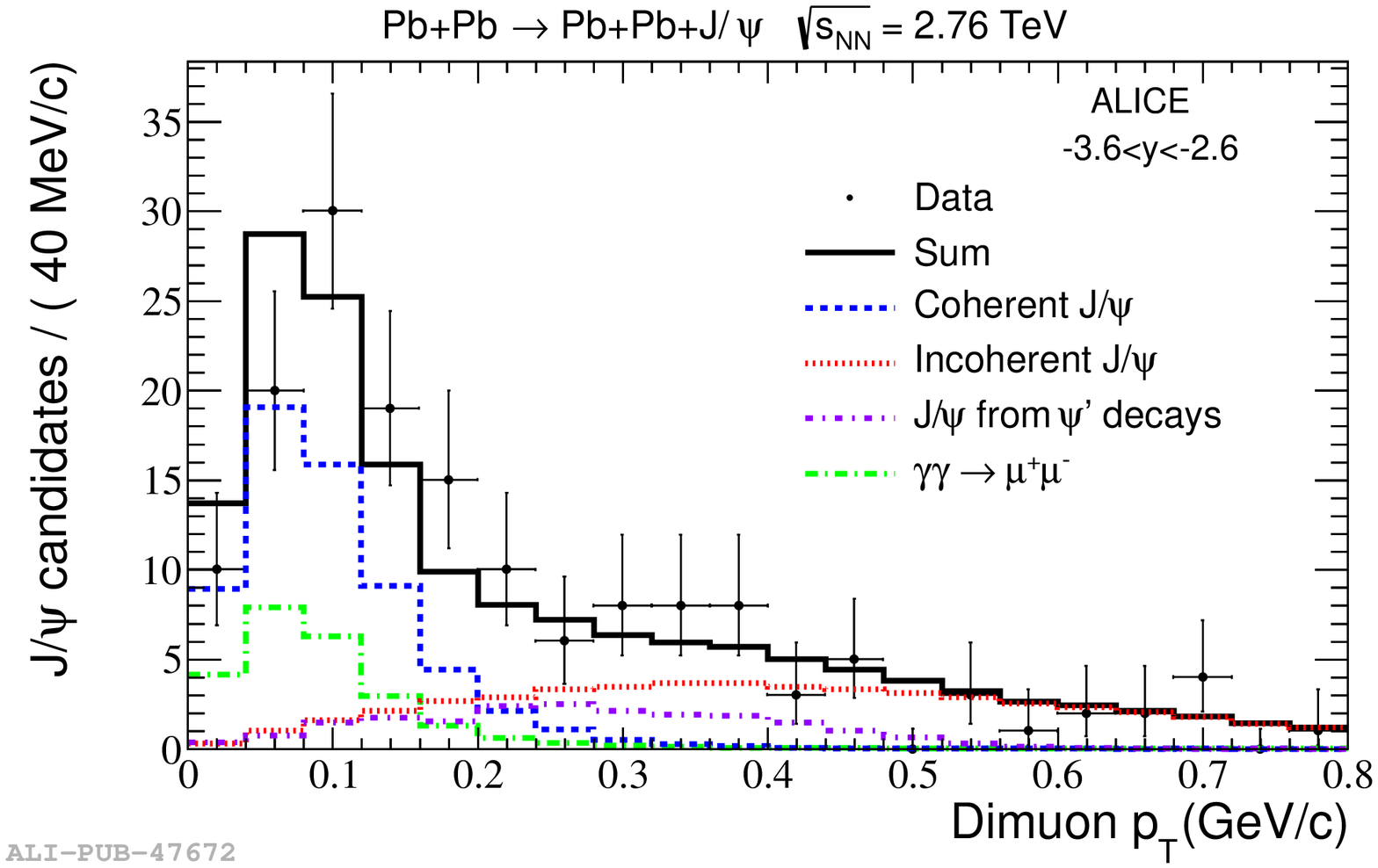}
}
\caption{Di-muon invariant mass~(left) and $p_{\rm T}$~(right) distributions in ultra-peripheral collisions at $\sqrt{s_{\rm NN}}$=2.76 TeV.}  
\end{figure}
\subsection{ Mid-rapidity analysis}
The $\jpsi$ production in UPC at mid-rapidity was measured using a sample of
6.5$\cdot 10^6$ events collected during the 2011 Pb-Pb run. The trigger required a number of
fired pad-OR in the time-of-flight detector (TOF) in the range 2$\leq$N$\leq$ 6, with at least two of
them back-to-back, two hits in the silicon pixel detector (SPD) and vetoes on VZERO counters
at forward and backward rapidity. Track reconstruction was performed with the time projection
chamber (TPC) and the silicon tracking system in about two pseudorapidity units. Events with
only two opposite sign good-quality tracks coming from a reconstructed primary vertex have
been selected. $\jpsi$ signal was extracted both in di-electron and di-muon channels, which were
separated by the energy deposition in TPC. A sample was enriched with coherent events by
applying a cut $p_{\rm T}<$200~MeV/$c$ for di-muons ($p_{\rm T}<$300~MeV/$c$ for di-electrons) and requiring a
ZDC energy deposit $E$$<$6 TeV. 
The number of coherent $\jpsi$ was then obtained using the same prescription described in the forward
analysis, giving 255$\pm$16(stat)$^{+14}_{-13}$(syst) 
for di-muon channel and 212$\pm$ 32(stat)$^{+14}_{-13}$(syst) 
for di-electron channel, respectively. 
Cross section normalization was performed with respect to the integrated luminosity
which was measured using a trigger for the most central hadronic Pb-Pb collisions.
Results have been recently published in~\cite{ref9}, giving for $|y|<$ 0.9 a coherent $\jpsi$ cross section 
${\rm d}\sigma_{coh}^{\jpsi}/{\rm d}y$= $2.38^{+0.34}_{-0.24}$(stat+syst) mb.
\subsection{Comparison with models}
The measured cross section can be compared to the available models~\cite{ref2,ref3,ref4,ref5,ref6,ref7}. The
models by Lappi and Mantysaari~\cite{ref3}, Cizek et al.~\cite{ref4}, and Goncalves and Machado~\cite{ref5} are based
on the Color Dipole Model (CDM). The model by Klein and Nystrand~\cite{ref7}, which is incorporated
in STARLIGHT~\cite{ref10}, uses data from exclusive vector meson production at HERA as input to a
Glauber calculation of the cross section for nuclear targets. The models by Rebyakova et al~\cite{ref2}
and Adeluyi and Bertulani~\cite{ref6}, calculate the cross section directly from the nuclear gluon
distribution with the forward scattering amplitude being proportional to the gluon distribution
squared. Rebyakova et al. calculate the modifications to the nuclear gluon distribution in the
leading twist approximation, while Adeluyi and Bertulani use standard parameterizations
(HKN07, EPS09, and EPS08). 
The results are shown in figure 2~(left). One can see that models which are based on
the color dipole model generally give a higher cross section than those which calculate the cross
section directly from the gluon distribution. 
Best agreement is found for models which include nuclear gluon shadowing
consistent with the EPS09 parameterizations. 
\begin{figure}[thb]
\mbox
{
\includegraphics[width=7.2cm,keepaspectratio]{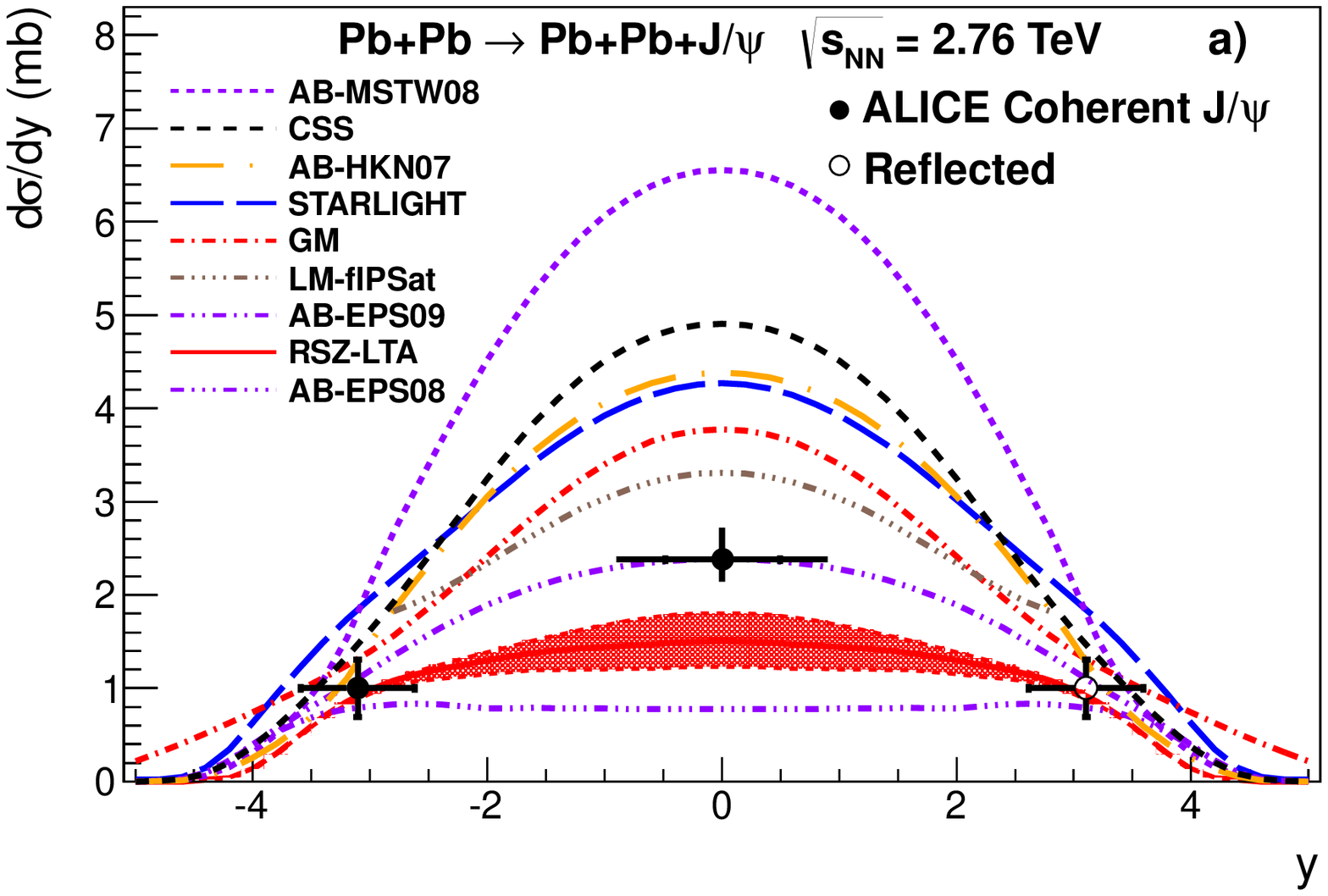}
\includegraphics[width=7.2cm,keepaspectratio]{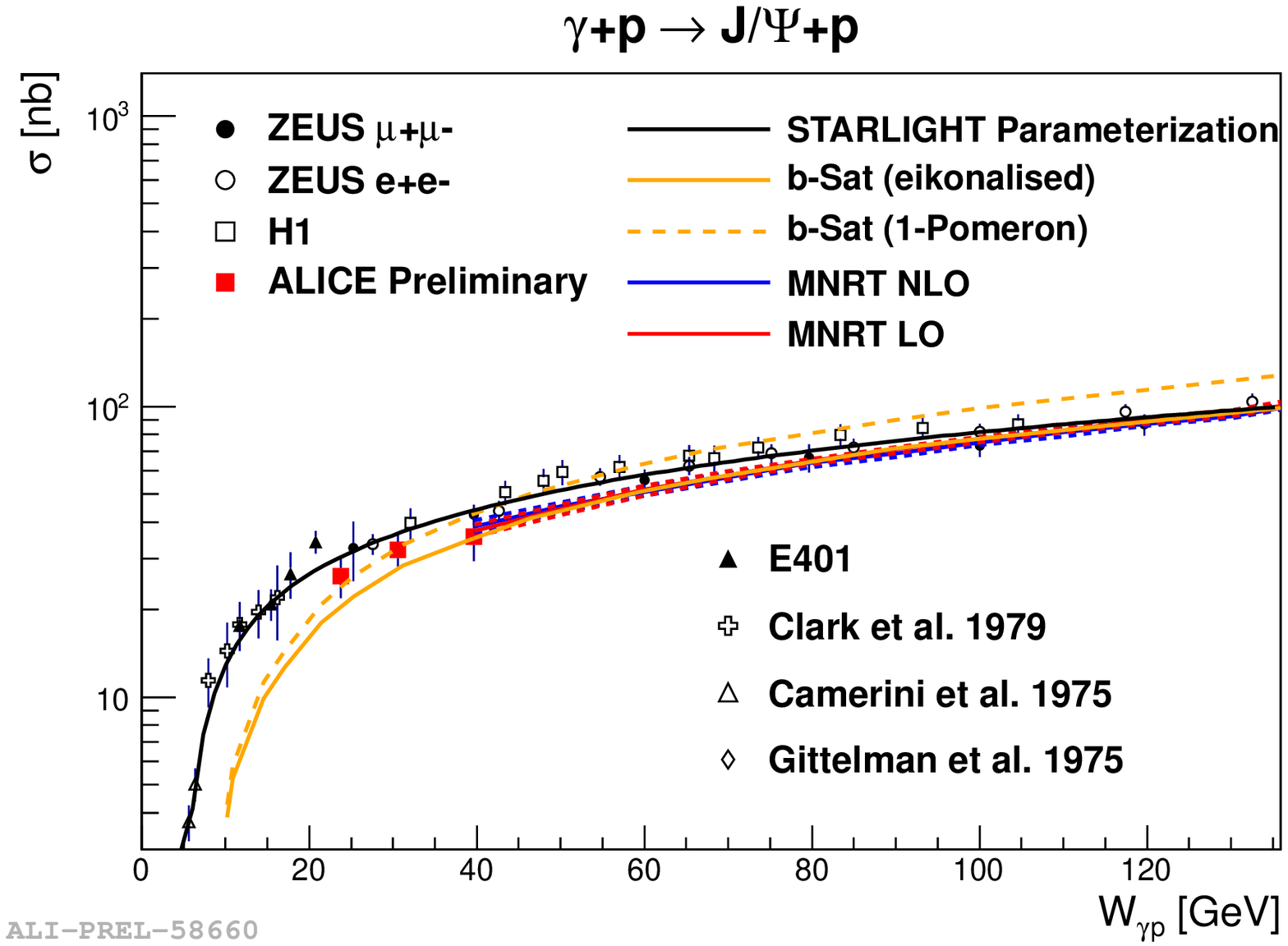}
}
\caption{Left: Pb-Pb coherent $\jpsi $~cross section compared to models in Pb-Pb collisions. Right: 
$\gamma$+p $\rightarrow$$\jpsi$+p cross section in p-Pb collisions compared with data from previous experiments.} 
\label{FigureLabel2}
\end{figure}
\section{$\jpsi$ production in p-Pb collisions}
Photons produced by a high energy nucleus colliding with a proton, offer the 
possibility to shed light on the proton structure functions. ALICE can measure
the p+Pb$\rightarrow$p+Pb+$\jpsi$ cross section in a wide range of rapidities, 
using forward~(two leptons in the muon arm), semi-forward~(one lepton in the muon arm 
and another one in the barrel) and midrapidity~(two leptons in the barrel) events.
Moreover by reversing the proton and Pb nuclei direction, the 2013 p-Pb run offered the possibility
to extend further the rapidity interval.  
Since the rapidity is connected to the $\gamma$p center of mass energy, $W_{\gamma p}$, these 
categories of events allow to study the cross section in the interval 20$<W_{\gamma p}<$950 GeV.
The  $\gamma$+p $\rightarrow$$\jpsi$+p 
cross section can be easily obtained by the  p+Pb$\rightarrow$p+Pb+$\jpsi$ 
cross section, dividing by the photon spectrum emitted by the Pb nucleus, computed by several Monte Carlo simulations  
and constrained by a recent ALICE analysis~\cite{ref9}.
The study of this cross section is one of the key tools to search for gluon saturation at small Bjorken-$x$.
As a first step ALICE focused on the low energy events ($W_{\gamma p}$$\simeq$10-40 GeV) to check the compatibility
between the ALICE measurements
with the previous measurements from HERA. Figure 2~(right) shows the three experimental points obtained by ALICE, fully compatible with the 
previous one.
The next step will be the analysis of the highest energy events, going beyond the 
region already explored by HERA.  
%
%
%
%
\begin{footnotesize}

\end{footnotesize}
\end{document}